\documentclass[letterpaper, 10 pt, conference]{ieeeconf}  

\IEEEoverridecommandlockouts                              

\newcommand{\acron}{ZETA}
\newcommand{\acrone}{Zonotope-based EsTimation and fAult diagnosis of discrete-time systems}

\title{\LARGE \bf {\acron}: a library for {\acrone}
}

\author{Brenner S. Rego$^{1}$, Joseph K. Scott$^{2}$, Davide M. Raimondo$^{3}$, Marco H. Terra$^{1}$, and Guilherme V. Raffo$^{4}$
\thanks{*This work was partially supported by the Brazilian agencies CNPq, under grants %
317058/2023-1 and 422143/2023-5; FAPESP, under the grant %
2022/05052-8; and CAPES through the Academic Excellence Program (PROEX).}
\thanks{$^{1}$Brenner S. Rego and Marco H. Terra are with the Department of Electrical and Computer Engineering, University of São Paulo, São Carlos, SP 13566-590, Brazil. {\tt\small brennersr7@usp.br, terra@sc.usp.br}
        }%
\thanks{$^{2}$Joseph K. Scott is with the Department of Chemical and Biomolecular Engineering, Georgia Institute of Technology, 311 Ferst Dr., Atlanta, 30318, GA, USA. {\tt\small joseph.scott@chbe.gatech.edu}
        }%
\thanks{$^{3}$Davide M. Raimondo is with the Department of Engineering and Architecture, University of Trieste, 34127, Trieste, Italy. {\tt\small davidemartino.raimondo@dia.units.it}
}
\thanks{$^{4}$Guilherme V. Raffo is with the Department of Electronics Engineering and the Graduate Program in Electrical Engineering, Federal University of Minas Gerais, Belo Horizonte, MG 31270-901, Brazil. {\tt\small raffo@ufmg.br}
        }%
}

\usepackage[american]{babel}
\usepackage[utf8]{inputenc}
\usepackage[T1]{fontenc}
\usepackage{graphicx}
\usepackage{amsmath}
\usepackage{amssymb}
\usepackage{amsfonts}
\usepackage{bm}
\usepackage{hhline}
\usepackage{csquotes}
\usepackage{xcolor}
\usepackage{import}
\usepackage{algorithm}
\usepackage{algpseudocode}
\usepackage{dsfont}
\usepackage{multirow}
\usepackage{url}

\usepackage{soul}

\usepackage{tablefootnote}

\usepackage{listings}

\newtheorem{definition}{\bf{Definition}}
\newtheorem{remark}{\bf{Remark}}

\allowdisplaybreaks 

\newcommand{\mbf}[1]{\ensuremath{{\mathbf{#1}}}}

\newcommand{\half}{\ensuremath{\frac{1}{2}}}

\newcommand{\eye}[1]{\ensuremath{\mbf{I}_{#1}}}

\newcommand{\zeros}[2]{\ensuremath{\bm{0}_{#1\times#2}}}
\newcommand{\ones}[2]{\ensuremath{\bm{1}_{#1\times#2}}}

\newcommand{\real}[1]{\ensuremath{\text{Re}(#1)}}

\newcommand{\realset}{\ensuremath{\mathbb{R}}}
\newcommand{\realsetmat}[2]{\ensuremath{\mathbb{R}^{#1\times#2}}}
\newcommand{\intvalset}{\ensuremath{\mathbb{I}\mathbb{R}}}

\newcommand{\lbound}{\ensuremath{\text{L}}}
\newcommand{\ubound}{\ensuremath{\text{U}}}

\newcommand{\lb}[1]{{#1}^\lbound}
\newcommand{\ub}[1]{{#1}^\ubound}

\newcommand{\poly}{\ensuremath{_\text{P}}}
\newcommand{\zon}{\ensuremath{_\text{Z}}}
\newcommand{\czon}{\ensuremath{_\text{CZ}}}
\newcommand{\lzon}{\ensuremath{_\text{LZ}}}
\newcommand{\strip}{\ensuremath{_\text{S}}}

\newcommand{\intval}[1]{\ensuremath{[\lb{#1}, \ub{#1}]}}

\newcommand{\noarg}{\ensuremath{\_\,}}

\newcommand{\code}[1]{\texttt{\detokenize{#1}}}

\newcommand{\bibfolder}{Bibliography}

\begin{document}

\maketitle
\thispagestyle{empty}
\pagestyle{empty}

\begin{abstract}
This paper introduces {\acron}, a new MATLAB library for {\acrone}. It features user-friendly implementations of set representations based on zonotopes, namely zonotopes, constrained zonotopes, and line zonotopes, in addition to a basic implementation of interval arithmetic. This library has capabilities starting from the basic set operations with these sets, including propagations through nonlinear functions using various approximation methods. The features of {\acron} allow for reachability analysis and state estimation of discrete-time linear, nonlinear, and descriptor systems, in addition to active fault diagnosis of linear systems. Efficient order reduction methods are also implemented for the respective set representations. Some examples are presented in order to illustrate the functionalities of the new library.
\end{abstract}


\section{Introduction}

In the last decades, many areas of scientific investigation have started to use set-based algorithms for different goals. From reachability analysis of dynamic systems \cite{Althoff2021Reach,Yang2023Guaranteed,bird2023hybrid,alanwar2023data}, to state \cite{Alamo2005a,raissi2011interval} and parameter estimation \cite{Rego2022Joint}, fault diagnosis \cite{Raimondo2016,Zhang2023}, invariant \cite{schafer2023scalable} and controllable sets \cite{vinod2025projection}, and model predictive control \cite{Nascimento2023,Richard2024MPCCZ}, set-based operations have gained attention because they allow to generate guaranteed enclosures for  dynamic systems subject to bounded uncertainties.

Thanks to their advantages for important set operations such as the Minkoswki sum and linear image, zonotopes have been used to obtain tight enclosures for the trajectories of discrete-time linear systems \cite{Combastel2003,Serry2022}. Such enclosures have been successfully used in accurate state estimation and active fault diagnosis \cite{Scott2014}. Constrained zonotopes (CZs), an extension of zonotopes proposed in \cite{Scott2016}, allowed further improvements in these topics \cite{Raimondo2016}, mainly due to their capability of representing arbitrary convex polytopes. CZs retain key computational advantages of zonotopes, including efficient complexity reduction algorithms. Constrained zonotopes were instrumental in achieving efficient  state estimation and active fault diagnosis of discrete-time linear descriptor systems \cite{Rego2020b}, whose trajectories have both a dynamic and static nature. To overcome the limitations of CZs in representing unbounded sets, \cite{Rego2024LineZonArxiv} introduced line zonotopes (LZs), an extension of the original framework that retains support for effective order reduction methods. %

Zonotopic sets have been especially important for the development of computationally affordable reachability analysis and state estimation algorithms for discrete-time nonlinear systems \cite{dePaula2022}. Zonotope methods have been proposed since the 90's starting with a Mean Value Theorem approach \cite{Kuhn1998}. This has been extended in \cite{Alamo2005a}, and a Taylor's Theorem approach has been proposed in \cite{Combastel2005}. These two algorithms have been improved by using constrained zonotopes in \cite{Rego2020}, and further extended for nonlinear measurement equations and invariants in \cite{Rego2021}. Difference of convex (DC) programming principles have been used in \cite{Alamo2008} and \cite{DePaula2024DC} for state estimation using zonotopes and CZs, respectively. Finally, polyhedral relaxation techniques have been used in \cite{Rego2024CZPRCDC} and \cite{Rego2025PolyRelaxArxiv} to obtain improved CZ enclosures for discrete-time nonlinear systems.

The main objective of this paper is to introduce {\acron}, a new MATLAB library for {\acrone}\footnote{See \url{https://github.com/Guiraffo/ZETA-releases}.}. Thanks to their computational advantages, zonotopic sets are very important for the development of set-based algorithms. However, the underlying concepts for such algorithms can be difficult to tackle by the general audience, while robust and reliable implementations are not straightforward. A few libraries are available in different languages \cite{Althoff2015CORA,Koeln2024zonolab,Vinod2024Pycvxset}, providing algorithms for basic operations and reachability analysis using zonotopic sets. However, robust implementations of several of the mentioned methods are still not available as off-the-shelf algorithms, especially the polyhedral relaxation techniques proposed in \cite{Rego2024CZPRCDC,Rego2025PolyRelaxArxiv}, active fault diagnosis methodologies \cite{Scott2014}, and line zonotopes \cite{Rego2024LineZonArxiv}. The aim of {\acron} is to fulfill this relevant gap, including (besides various zonotopic set representations and their basic operations): (i) efficient implementations of CZ complexity reduction algorithms, (ii) state estimation and fault diagnosis of discrete-time linear systems, (iii) several approximation methods for enclosing the trajectories of discrete-time nonlinear systems, (iv) nonlinear state estimation, and (v) active fault diagnosis methods.

\subsection*{Notation}

Lowercase italic letters denote scalars, lowercase bold letters denote vectors, uppercase bold letters denote matrices, and uppercase italic letters denote general sets.  Moreover, $\zeros{n}{m}$ and $\ones{n}{m}$ denote $n \times m$ matrices of zeros and ones, respectively. The $n \times n $ identity matrix is denoted by $\eye{n}$. The set of real numbers is denoted by $\realset$. Additionally, let $\bm{\alpha} : \realset^{n_s} \to \realset^{n_\alpha}$ be a nonlinear function with argument $\mbf{s}$. Functions with set-valued arguments $S$ satisfying $\mbf{s} \in S$ denote the exact image of the set under the function, i.e., $\bm{\alpha}(S)\triangleq\{\bm{\alpha}(\mbf{s}): \mbf{s}\in S\}$. 

\section{Mathematical background} \label{sec:background}

Consider a real matrix $\mbf{R} \in \realset^{m \times n}$ and sets $Z, W \subset \realset^{n}$, $Y \subset \realset^{m}$. The linear mapping, Minkowski sum, generalized intersection, and the Cartesian product are defined as
\begin{align}
\mbf{R}Z & \triangleq \{ \mbf{R} \mbf{z} : \mbf{z} \in Z\}, \label{eq:pre_limage}\\
Z \oplus W & \triangleq \{ \mbf{z} + \mbf{w} : \mbf{z} \in Z,\, \mbf{w} \in W\}, \label{eq:pre_msum}\\
Z \cap_{\mbf{R}} Y & \triangleq \{ \mbf{z} \in Z : \mbf{R} \mbf{z} \in Y\}, \label{eq:pre_intersection}\\
Z \times Y & \triangleq \{ (\mbf{z},\mbf{y}) : \mbf{z} \in Z,\, \mbf{y} \in Y\}, \label{eq:pre_cartprod}
\end{align}
respectively. Moreover, $\text{conv}(Z,W)$ denotes the convex hull of $Z$ and $W$. 

To accomplish its various capabilities, {\acron} implements a few different set representations: intervals, strips, zonotopes, constrained zonotopes, line zonotopes, and convex polytopes in halfspace representation. These are defined below.

\begin{definition} \rm \label{def:intervals}
	Let $\intvalset^n$ denote the set of all non-empty compact intervals in $\realset^n$. If $\lb{\mbf{x}}, \ub{\mbf{x}} \in \realset^n$ with $\lb{\mbf{x}} \leq \ub{\mbf{x}}$, an \emph{interval} $X \in \intvalset^n$ is defined as $X \triangleq \{\mbf{x} \in \realset^n: \lb{\mbf{x}} \leq \mbf{x} \leq \ub{\mbf{x}}\} \triangleq \intval{\mbf{x}}$.
\end{definition}

\begin{definition} \rm \label{def:strips}
	  A set $S \subset \realset^{n_s}$ is a \emph{strip} if there exists a tuple $(\mbf{p}_s,d_s,\sigma_s) \in \realset^{n_s} \times \realset \times \realset$ such that $S = \{ \mbf{s} \in \realset^{n_s} : |\mbf{p}_s^T \mbf{s} - d_s| \leq \sigma_s\}$. We use the shorthand notation $(\mbf{p}_s,d_s,\sigma_s)\strip$ for strips.
\end{definition}

\begin{definition} \rm \label{def:zonotopes}
	A set $Z \subset \realset^{n_z}$ is a \emph{zonotope} with $n_g$ generators if there exists a tuple $(\mbf{G}_z,\mbf{c}_z) \in \realsetmat{n_z}{n_g} \times \realset^n$ such that $Z = \left\{ \mbf{c}_z + \mbf{G}_z \bm{\xi} : \| \bm{\xi} \|_\infty \leq 1\right\}$. We use the shorthand notation $(\mbf{G}_z,\mbf{c}_z)\zon$ for zonotopes.
\end{definition}

\begin{definition} \cite{Scott2016} \rm \label{def:czonotopes}
	A set $Z \subset \realset^{n_z}$ is a \emph{constrained zonotope} with $n_g$ generators and $n_c$ constraints if there exists a tuple $(\mbf{G}_z,\mbf{c}_z,\mbf{A}_z,\mbf{b}_z) \in \realsetmat{n}{n_g} \times \realset^n \times \realsetmat{n_c}{n_g} \times \realset^{n_c}$ such that $Z = \left\{ \mbf{c}_z + \mbf{G}_z \bm{\xi} : \| \bm{\xi} \|_\infty \leq 1, \mbf{A}_z \bm{\xi} = \mbf{b}_z \right\}$. We use the shorthand notation $(\mbf{G}_z,\mbf{c}_z,\mbf{A}_z,\mbf{b}_z)\czon$ for constrained zonotopes.
\end{definition}

\begin{definition} \rm \label{def:lzonotopes}
	A set $Z \subseteq \realset^{n_z}$ is a \emph{line zonotope} with $n_\ell$ lines, $n_g$ generators, and $n_c$ constraints, if there exists a tuple $(\mbf{M}_z,\mbf{G}_z,\mbf{c}_z,\mbf{S}_z,\mbf{A}_z,\mbf{b}_z) \in \realsetmat{n_z}{n_\ell} \times \realsetmat{n_z}{n_g} \times \realset^n \times \realsetmat{n_c}{n_\ell} \times \realsetmat{n_c}{n_g} \times \realset^{n_c}$ such that $Z = \{\mbf{M}_z \bm{\delta} + \mbf{G}_z \bm{\xi} + \mbf{c}_z:  \bm{\delta} \in \realset^{n_\ell}, \| \bm{\xi} \|_\infty \leq 1, \mbf{S}_z \bm{\delta} + \mbf{A}_z \bm{\xi} = \mbf{b}_z \}.$ We use the shorthand notations $(\mbf{M}_z,\mbf{G}_z,\mbf{c}_z,\mbf{S}_z,\mbf{A}_z,\mbf{b}_z)\lzon$ and $(\mbf{M}_z,\mbf{G}_z,\mbf{c}_z)\lzon$ for line zonotopes with and without constraints, respectively.
\end{definition}

\begin{definition} \rm \label{def:hrep}
	A set $P \subset \realset^n$ is a \emph{convex polytope} in halfspace representation if there exists $(\mbf{H}_p, \mbf{k}_p, \mbf{A}_p, \mbf{b}_p) \in \realsetmat{n_h}{n} \times \realset^{n_h} \times \realsetmat{n_{c_p}}{n} \times \realset^{n_{c_p}}$ such that $P = \{ \mbf{x} \in \realset^n : \mbf{H}_p \mbf{x} \leq \mbf{k}_p, \ \mbf{A}_p \mbf{x} = \mbf{b}_p \}$. We use the shorthand notation $(\mbf{H}_p,\mbf{k}_p,\mbf{A}_p,\mbf{b}_p)\poly$ for convex polytopes.
\end{definition}

\section{Core features of {\acron}} \label{sec:core}

The {\acron} toolbox makes extensive use of Object Oriented Programming (OOP) and operator overloading to implement set computations. Auxiliary methods necessary for set-based state estimation and active fault diagnosis are also present. This section describes the core features of the library.

\subsection{Structure, dependencies, and external libraries}

ZETA files are organized according to the following folder structure:
\begin{itemize}
    \item \code{packages}: contains utility algorithms used for internal computations.%
    \item \code{objects}: contains implementations of classes (set representations and other objects) and respective methods.
    \item \code{estimation}: contains implementations of state estimation methods.
    \item \code{faultdiag}: contains implementations of fault diagnosis methods.
    \item \code{demos}: contains numerical examples that illustrate the capabilities of {\acron}.
\end{itemize}

Currently, the minimum requirements for {\acron} to run properly in a MATLAB installation are YALMIP \cite{Lofberg2004} and the MATLAB Optimization Toolbox. For improved results in optimization problems and allowing the solution of mixed-integer programs, {\acron} supports Gurobi \cite{gurobi}. Plotting capabilities are extended if MPT (Multi-parametric toolbox) \cite{Herceg2013} is available.

\subsection{Class constructors for sets}%

{\acron} implements a few classes for its core functionalities with different set representations: \code{Interval}, \code{Strip}, \code{Zonotope}, \code{CZonotope}, and \code{LZonotope}. This paper describes briefly each one of these classes and their methods. A full description can be found by using the \code{help} command at the respective functions.

The \code{Interval} class implements intervals as in Definition \ref{def:intervals}. An object of the \code{Interval} class stores the interval endpoints as properties. An \code{Interval} object can be created through different syntaxes: (i) \code{Interval} returns the scalar degenerate interval $[0,0]$; (ii) \code{Interval(a)} for double \code{a} creates a degenerate interval $[a,a$]; and (iii) \code{Interval(a,b)} for doubles \code{a} and \code{b} (with $a \leq b$), creates the interval $[a,b]$. Interval vectors and matrices can be created using consistent vector and matrix arguments, respectively.

The \code{Strip} class implements strips as in Definition \ref{def:strips}. For double scalars \code{d} and \code{s}, and a double vector \code{p}, a \code{Strip} object can be created through different syntaxes: (i) \code{Strip(p)} creates the strip $(\mbf{p},0,1)\strip$; (ii) \code{Strip(p,d)} creates the strip $(\mbf{p},d,1)\strip$; and (iii) \code{Strip(p,d,s)} creates the strip $(\mbf{p},d,s)\strip$.

\begin{table*}[!htb]
	\scriptsize
	\centering
	\caption{Methods for creating zonotopic sets in {\acron}}
	\begin{tabular}{l l | l l | l l} \hline
		Zonotopes & Result & CZs & Result & LZs & Result \\ \hline
		\code{Zonotope(c)} & $(\noarg,\mbf{c})\zon$ & \code{CZonotope(c)} & $(\noarg,\mbf{c},\noarg,\noarg)\czon$ & \code{LZonotope(c)} & $(\noarg,\mbf{c},\noarg)\lzon$ \\
		\code{Zonotope(c,G)} & $(\mbf{G},\mbf{c})\zon$ & \code{CZonotope(c,G)} & $(\mbf{G},\mbf{c},\noarg,\noarg)\czon$ & \code{LZonotope(c,G)} & $(\noarg,\mbf{G},\mbf{c})\lzon$\\
        & & \code{CZonotope(c,G,A,b)} & $(\mbf{G},\mbf{c},\mbf{A},\mbf{b})\czon$ & \code{LZonotope(c,G,M)} & $(\mbf{M},\mbf{G},\mbf{c})\lzon$\\
        & & & & \code{LZonotope(c,G,A,b)} & $(\noarg,\mbf{G},\mbf{c},\noarg,\mbf{A},\mbf{b})\lzon$\\    
        & & & & \code{LZonotope(c,G,M,S,A,b)} & $(\mbf{M},\mbf{G},\mbf{c},\mbf{S},\mbf{A},\mbf{b})\lzon$\\            
        & & & & \code{LZonotope.realset(n)} & $(\eye{n},\noarg,\zeros{n}{1})\lzon$\\                    
		\hline 
	\end{tabular} \normalsize
	\label{tab:setconstructors}
\end{table*}

The \code{Zonotope}, \code{CZonotope}, and \code{LZonotope} classes implement zonotopes as in Definition \ref{def:zonotopes}, constrained zonotopes as in Definition \ref{def:czonotopes}, and line zonotopes as in Definition \ref{def:lzonotopes}, respectively. We refer to these set representations as zonotopic sets. For real matrices \code{G}, \code{A}, \code{M}, \code{S}, vectors \code{c}, \code{b}, and a natural number \code{n}, Table \ref{tab:setconstructors} shows the main ways of creating objects of the different zonotopic set classes. In the notation show in Table \ref{tab:setconstructors}, `$\noarg$' denotes an empty argument with appropriate dimensions. For instance, $(\noarg,\mbf{c})\zon$ is a degenerated zonotope containing only the vector $\mbf{c}$.

\begin{figure}[!thb]
	\centering{
		\def\svgwidth{\columnwidth}
  {\tiny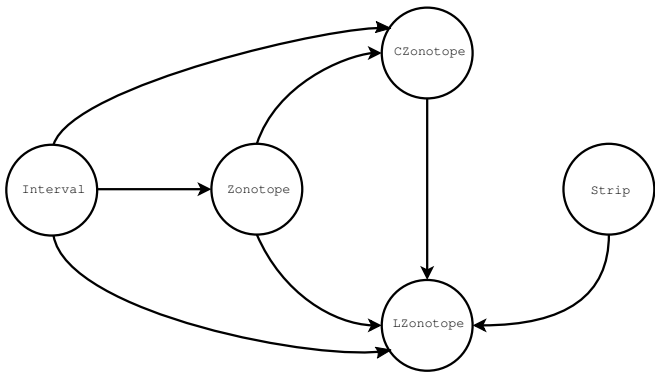}
		\caption{Conversions between set representations implemented in {\acron}.}\label{fig:classconstructors}}%
\end{figure}

The respective class constructors implement several conversions between set representations in {\acron}. Fig.~\ref{fig:classconstructors} shows the conversions allowed. For instance, if \code{B} is an \code{Interval} object, then \code{CZonotope(B)} converts \code{B} into an object of the \code{CZonotope} class using the appropriate formula. Conversions into vertex and/or half-space representations are also implemented when possible. Finally, each one of the set representations implemented in {\acron} has its own \code{plot} method.

\subsection{Interval arithmetic}
Basic interval arithmetic routines are implemented in {\acron} as methods of the \code{Interval} class through operator overloading. To improve efficiency, the interval arithmetic routines in our library do not implement outward rounding\footnote{Future versions of {\acron} will have optional support to INTLAB \cite{Rump1999} for interval arithmetic with outward rounding.}. Table \ref{tab:intervaloperations} shows many interval operations included in {\acron}\footnote{The sampling methods in {\acron} allow the user to choose from different distributions (default is uniform).}.
\begin{table}[htbp!]
	\scriptsize
	\centering
	\caption{Operations with \code{Interval} objects in {\acron}}
	\begin{tabular}{l l} \hline
		Method & Operation \\ \hline
        \code{mid(a)}, \code{rad(a)}, \code{diam(a)} & $\half(\ub{a} {+} \lb{a})$, $\half(\ub{a} {-} \lb{a})$, $\ub{a} {-} \lb{a}$ \\
        \code{intersect(a,b)}, \code{hull(a,b)} & $a \cap b$, $[\text{min}(\lb{a},\lb{b}),\text{max}(\ub{a},\ub{b})]$ \\
        \code{a + b}, \code{a - b}, \code{a*b}, \code{a/b} & $a+b$, $a-b$, $ab$, $\frac{a}{b}$ \\
        \code{a^n}, \code{sqrt(a)}, \code{exp(a)}, \code{log(a)} & $a^n$, $\sqrt{a}$, $e^a$, $\ln(a)$ \\
		\code{sin(a)}, \code{cos(a)}, \code{tan(a)}, \code{abs(a)} & $\sin(a)$, $\cos(a)$, $\tan(a)$, $|a|$ \\
        \code{norm(a,1)}, \code{norm(a,2)} & $\|\mbf{a}\|_1$, $\|\mbf{a}\|_2$ \\
        \code{sample(a,n)} & Generates $n$ samples in $a$ \\
		\hline 
		Method & Returns true if \\ \hline
        \code{a>b}, \code{a>=b} & $\lb{a} > \ub{b}$, $\lb{a} \geq \ub{b}$ \\
        \code{a<b}, \code{a<=b} & $\ub{a} < \lb{b}$, $\ub{a} \leq \lb{b}$ \\
        \code{isinside(a,x)} & $x \in a$ \\
		\hline         
	\end{tabular} \normalsize
	\label{tab:intervaloperations}
\end{table}

\subsection{Operations with zonotopes and extensions}

Several routines with zonotopes, CZs, and LZs are implemented in {\acron}. Tables \ref{tab:zonoperations}, \ref{tab:czonoperations}, and \ref{tab:lzonoperations} show the basic operations available for \code{Zonotope}, \code{CZonotope}, and \code{LZonotope} objects\footnote{For a \code{LZonotope} $Z$, \code{reduction} also eliminates all the removable lines of $Z$.}, respectively. The usage of these operations is demonstrated in various example files available in the \code{demos/basic} folder.

\begin{figure}[!thb]
	\centering{
		\def\svgwidth{\columnwidth}
  {\scriptsize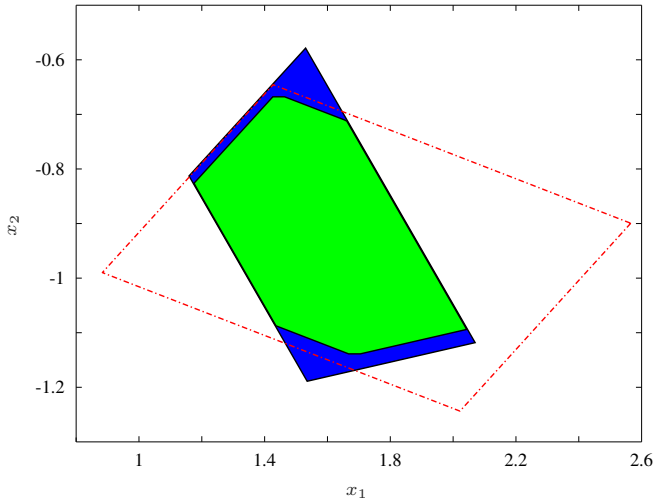}
		\caption{Constrained zonotope $Z$ to be reduced (solid green), and the reduced $\bar{Z}$ obtained by {\acron} (solid blue) and CORA (dot-dashed red). The \code{CZonotope} objects have been plotted using \code{plot}.}\label{fig:reductioncomparison}}
\end{figure}

To illustrate one of the main advantages of our library, we present a CZ complexity reduction example, available in \code{demos/basic/CZonotope/demo_reduction.m}. This example consists of reducing a constrained zonotope $Z \subset \realset^2$ with 47 generators and 15 constraints to another constrained zonotope, $\bar{Z} \subset \realset^2$, with four generators and two constraints. In {\acron}, this is accomplished by \code{reduction(Z,4,2)}, which implements the reduction methods proposed in \cite{Scott2016}. Fig.~\ref{fig:reductioncomparison} shows the original set $Z$ (solid green) along with $\bar{Z}$ (solid blue). For comparison purposes, we also compute the same operation using CORA \cite{Althoff2015CORA} version 2025.1.0 (dot-dashed red), using the `scott' method for generator reduction. As can be noticed, the reduced enclosures $\bar{Z}$ obtained by {\acron} and CORA are very distinct for this example, with the enclosure provided by {\acron} being less conservative.

\begin{table}[!htbp]
	\scriptsize
	\centering
	\caption{Operations with \code{Zonotope} objects in {\acron}}
	\begin{tabular}{l l} \hline
		Method & Operation \\ \hline
        \code{[Z,W]}, \code{[Z;W]} & $Z \times W$ \\
        \code{Z + W}, \code{Z - W}, \code{R*Z} &  $Z \oplus W$, $Z \oplus (-W)$, $\mbf{R}Z$ \\
        \code{intersection(Z,S)} & Zonotope enclosing $Z \cap S$ ($S$ is a \code{Strip} object)\\
        \code{convhull(Z,W)} &  $\text{conv}(Z,W)$ (Theorem 5 in \cite{Raghuraman2022})\\
        \code{intervalhull(Z)} &  $\square Z$ (Remark 3 in \cite{Kuhn1998}) \\
        \code{unlift(Z)} & Unlifts $Z$ (if it is a lifted representation of a CZ) \\        
        \code{partopebound(Z)} & Returns a parallelotope enclosure of $Z$\\ 
        \code{permute(Z,...)} & Permutes the variables of $Z$\\ 
        \code{projection(Z,...)} & Projects $Z$ onto desired dimensions \\
        \code{radius(Z)} & Radius of $Z$ following different metrics \\
        \code{reduction(Z,ng)} & Reduces $Z$ to $n_g$ generators \\
        \code{sample(Z,n)} & Generates $n$ samples in $Z$ \\
        \code{volume(Z)} & Volume of $Z$ \\
        \code{hrep(Z)} & H-rep of $Z$ (Theorem 7 in \cite{Althoff2010}) \\
        \code{vrep(Z)} & V-rep of $Z$ \\
		\hline 
		Method & Returns true if \\ \hline
        \code{isinside(Z,z)} & $\mbf{z} \in Z$ (adaptation of Proposition 2 in \cite{Scott2016}) \\
		\hline         
	\end{tabular} \normalsize
	\label{tab:zonoperations}
\end{table}

\begin{table}[!htbp]
	\scriptsize
	\centering
	\caption{Operations with \code{CZonotope} objects in {\acron}}
	\begin{tabular}{l l} \hline
		Method & Operation \\ \hline
        \code{[Z,W]}, \code{[Z;W]} & $Z \times W$ \\
        \code{Z + W}, \code{Z - W}, \code{R*Z}, \code{Z & W} &  $Z \oplus W$, $Z \oplus (-W)$, $\mbf{R}Z$, $Z \cap W$ \\
        \code{intersection(Z,Y,R)} &  $Z \cap_\mbf{R} Y$\\
        \code{intersection(Z,H,k,A,b)} &  $Z \cap (\mbf{H},\mbf{k},\mbf{A},\mbf{b})\poly$ (Proposition 1 in \cite{Rego2024CZPRCDC}) \\
        \code{convhull(Z,W)} &  $\text{conv}(Z,W)$ (Theorem 5 in \cite{Raghuraman2022})\\
        \code{intervalhull(Z)} &  $\square Z$ (Property 1 in \cite{Rego2020}) \\
        \code{inclusion(Z,J)} & CZ-inclusion (Theorem 1 in \cite{Rego2020}) \\
        \code{closest(Z,h)} & Proposition 1 in \cite{Rego2020} \\
        \code{lift(Z)} & Lifts $Z$ as described in \cite{Scott2016} \\        
        \code{partopebound(Z)} & Returns a parallelotope enclosure of $Z$\\ 
        \code{permute(Z,...)} & Permutes the variables of $Z$\\ 
        \code{projection(Z,...)} & Projects $Z$ onto desired dimensions \\
        \code{radius(Z)} & Radius of $Z$ following different metrics \\
        \code{reduction(Z,ng,nc)} & Reduces $Z$ to $n_g$ gen. and $n_c$ cons. \\
        \code{rescale(Z)} & Rescales $Z$ using the methods in \cite{Scott2016} \\ 
        \code{sample(Z,n)} & Generates $n$ samples in $Z$ \\
        \code{volume(Z)} & Approximate volume of $Z$ \\
        \code{hrep(Z)} & H-rep of $Z$ (Theorem 1 in \cite{Scott2016}) \\
		\hline 
		Method & Returns true if \\ \hline
        \code{isempty(Z)} & $Z = \emptyset$ (Proposition 2 in \cite{Scott2016}) \\
        \code{isinside(Z,z)} & $\mbf{z} \in Z$ (Proposition 2 in \cite{Scott2016}) \\
		\hline         
	\end{tabular} \normalsize
	\label{tab:czonoperations}
\end{table}

\begin{table}[!htb]
	\scriptsize
	\centering
	\caption{Operations with \code{LZonotope} objects in {\acron}}
	\begin{tabular}{l l} \hline
		Method & Operation \\ \hline
        \code{[Z,W]}, \code{[Z;W]} & $Z \times W$ \\
        \code{Z + W}, \code{Z - W}, \code{R*Z} &  $Z \oplus W$, $Z \oplus (-W)$, $\mbf{R}Z$ \\
        \code{intersection(Z,Y,R)} &  $Z \cap_\mbf{R} Y$ \\
        \code{intervalhull(Z)} &  $\square Z$ ($Z$ must be bounded) \\        
        \code{projection(Z,...)} & Projects $Z$ onto desired dimensions \\
        \code{radius(Z)} & Radius of $Z$ following different metrics \\
        \code{reduction(Z,ng,nc)} & Reduces $Z$ to $n_g$ generators and $n_c$ constraints \\
		\hline 
		Method & Returns true if \\ \hline
        \code{isempty(Z)} & $Z = \emptyset$ (extended from Proposition 2 in \cite{Scott2016}) \\
        \code{isinside(Z,z)} & $\mbf{z} \in Z$ (extended from Proposition 2 in \cite{Scott2016}) \\
		\hline         
	\end{tabular} \normalsize
	\label{tab:lzonoperations}
\end{table}

\subsection{Polyhedral relaxations} \label{sec:polyrelax}

The \code{Polyrelax} class implements the polyhedral relaxation computations proposed in \cite{Rego2024CZPRCDC} and extended in \cite{Rego2025PolyRelaxArxiv}. Specifically, it implements the computation of the interval vector $Z$ and convex polytope $P_\varphi$ in Section 3 in \cite{Rego2025PolyRelaxArxiv}, along with the computation of the equivalent enclosure described in Section 4.1 of the same reference. For a factorable function $\bm{\varphi}(\mbf{s})$ with factors $\bm{\zeta}(\mbf{s})$ and input interval $S \in \intvalset^{n_s}$, both $Z$ and $P_\varphi$ contain the exact image of $\bm{\zeta}(\mbf{s})$ for $\mbf{s} \in S$. The projection of $P_\varphi$ onto the image of $\bm{\varphi}(\mbf{s})$ gives a polyhedral enclosure of the nonlinear function for $\mbf{s} \in S$. See \cite{Rego2024CZPRCDC}, \cite{Rego2025PolyRelaxArxiv} for details. 

Each \code{Polyrelax} object consists of a tuple of an \code{Interval} object and an integer index. It takes the \code{Interval} object as input, and the class constructor assigns an integer index. Our implementation relies mainly on the \code{Polyrelax} constructor, the indexing of each \code{Polyrelax} object, and static properties storing the interval vector $Z$ (static variable \code{Z}) and $P_\varphi$ (static variable \code{Hrep}), which are built procedurally. This will be illustrated in an example.

Let $\varphi(x,y)$ be a scalar nonlinear function. Let \code{Interval} objects \code{X} and \code{Y} be the domain intervals, and let \code{func(x,y)} denote the MATLAB function implementing $\varphi$. The script in Algorithm \ref{alg:polyrelaxexample} computes the interval $Z$ and polytope $P_\varphi$. The underlying computations (automated by {\acron}) in each step of this algorithm are explained below.

\paragraph{Step 1} \code{Polyrelax.clear} initializes the static variables \code{Z} and \code{Hrep} with empty values. This also allows the next \code{Polyrelax} object to have index $j=1$.

\paragraph{Steps 2, 3} \code{Polyrelax(X)} and \code{Polyrelax(Y)} create two \code{Polyrelax} objects, with \code{Interval} properties \code{X} and \code{Y}, and indices $j=1$ and $j=2$, respectively.

\paragraph{Step 4} \code{F = func(X_PR,Y_PR)} evaluates $\varphi(x,y)$ with \code{Polyrelax} objects as inputs. \code{F} is a \code{Polyrelax} object whose index is used for the projection of the polyhedral enclosure onto the image of $\varphi(x,y)$. Through operator overloading, for each $\alpha_j$ (as in Definition 2 in \cite{Rego2025PolyRelaxArxiv}), a new \code{Polyrelax} object is created, with
\begin{itemize}
    \item Index $j$ and interval property $Z_j$, computed using interval arithmetic of $\alpha_j$ and concatenated into the \code{Z} static variable; and
    \item Polyhedral relaxation $Q_j$, computed according to the respective operation in Section 3.1 and 3.2 in \cite{Rego2025PolyRelaxArxiv}, which is incorporated into the \code{Hrep} static variable through intersection.
\end{itemize}

\paragraph{Step 5} \code{Polyrelax.Z} retrieves the resulting interval vector $Z$.
\paragraph{Step 6} \code{Polyrelax.Hrep} retrieves the resulting polyhedral enclosure $P_\varphi$.

As it can be noticed, polyhedral relaxations of nonlinear functions can be computed in {\acron} effortlessly. Notably, indexing each existing \code{Polyrelax} object is essential to correctly constructing the interval vector $Z$ and each $Q_j$ by our implementation, which is completely automated in {\acron}. Table \ref{tab:polyrelaxoperations} illustrates the elementary functions implemented for \code{Polyrelax} objects \code{X} and \code{Y}.

\begin{table}[!htb]
	\scriptsize
	\centering
	\caption{Operations with \code{Polyrelax} objects in {\acron}}
	\begin{tabular}{l l} \hline
		Method & Operation \\ \hline
        \code{X + Y}, \code{X - Y}, \code{X*Y}, \code{X/Y} & $x+y$, $x-y$, $xy$, $\frac{x}{y}$ \\
        \code{X^n}, \code{sqrt(X)}, \code{exp(X)}, \code{log(X)} & $x^n$, $\sqrt{x}$, $e^x$, $\ln(x)$ \\
		\code{sin(X)}, \code{cos(X)}, \code{tan(X)}, \code{abs(X)} & $\sin(x)$, $\cos(a)$, $\tan(x)$, $|x|$ \\
        \code{norm(X,1)}, \code{norm(X,2)} & $\|\mbf{x}\|_1$, $\|\mbf{x}\|_2$ \\
		\hline 
	\end{tabular} \normalsize
	\label{tab:polyrelaxoperations}
\end{table}

\begin{algorithm}[!tb] 
	\caption{Computation of polyhedral relaxations for $\varphi(x,y)$ in {\acron} with input \code{Interval} objects \code{X} and \code{Y}.}
	\label{alg:polyrelaxexample}
	\scriptsize
	\begin{algorithmic}[1]
        \State \code{Polyrelax.clear;}
        \State \code{X_PR = Polyrelax(X);}
        \State \code{Y_PR = Polyrelax(Y);}
        \State \code{F = func(X_PR,Y_PR);}
        \State \code{Z = Polyrelax.Z;}
        \State \code{P = Polyrelax.Hrep;}
	\end{algorithmic}
	\normalsize
\end{algorithm}

\subsection{Discrete-time systems with bounded uncertainties} \label{sec:simulation}

The \code{DTsystem} class implements simulation routines. It enables a user-friendly interface for advanced features such as state estimation for a few classes of discrete-time systems subject to bounded uncertainties. 

The first class is a linear system described by
\begin{subequations} \label{eq:systemlinear}
\begin{align}
    \mbf{x}_k & = \mbf{A} \mbf{x}_{k-1} + \mbf{B}_w \mbf{w}_{k-1} + \mbf{B}_u \mbf{u}_{k-1}, \\
    \mbf{y}_k & = \mbf{C} \mbf{x}_{k} + \mbf{D}_v \mbf{v}_{k},
\end{align}
\end{subequations}
where $\mbf{x}_k \in \realset^{n_x}$ is the system state, $\mbf{w}_{k}$ is the process uncertainty, $\mbf{u}_k$ is the known input, $\mbf{y}_k$ is the measured output, and $\mbf{v}_k$ is the measurement uncertainty. The second class is a discrete-time linear descriptor system given by
\begin{subequations} \label{eq:systemdescriptor}
\begin{align}
    \mbf{E} \mbf{x}_k & = \mbf{A} \mbf{x}_{k-1} + \mbf{B}_w \mbf{w}_{k-1} + \mbf{B}_u \mbf{u}_{k-1}, \\
    \mbf{y}_k & = \mbf{C} \mbf{x}_{k} + \mbf{D}_v \mbf{v}_{k}.
\end{align}
\end{subequations}
The third class is a discrete-time system with nonlinear dynamics and nonlinear measurement equations, given by
\begin{subequations} \label{eq:systemnonlinear}
\begin{align}
    \mbf{x}_k & = \mbf{f}(\mbf{x}_{k-1},\mbf{w}_{k-1},\mbf{u}_{k-1}), \\
    \mbf{y}_k & = \mbf{g}(\mbf{x}_{k},\mbf{v}_{k}).
\end{align}
\end{subequations}

\code{DTSystem} objects can be easily created for these classes of discrete-time systems, as shown below:
\begin{itemize}
    \item \code{DTSystem('linear','A',A,'Bw',Bw,...)} creates a discrete-time linear system as in \eqref{eq:systemlinear}.%
    \item \code{DTSystem('descriptor','E',E,'A',A,...)} creates a discrete-time linear  descriptor system as in \eqref{eq:systemdescriptor}.%
    \item \code{DTSystem('nonlinear',modelname)} creates a discrete-time nonlinear  system as in \eqref{eq:systemnonlinear}, where \code{modelname} corresponds to the prefix of function names implementing $\mbf{f}$ and $\mbf{g}$.\footnote{For this syntax example, the current version of {\acron} requires that the user implements functions named \code{modelname_f.m} and \code{modelname_g.m} in the current working folder.}
\end{itemize}

A \code{DTSystem} object allows for straightforward simulation of the respective class of discrete-time system and plotting figures showing the resulting trajectory. Algorithm \ref{alg:dtsystemexample} illustrates some of the capabilities of a \code{DTSystem} object describing a linear system. Step 1 creates the object using the system matrices as inputs. Step 2 simulates the system for an initial state $\mbf{x}_0$, $N$ time steps, and uncertainties $\mbf{w}_k$ and $\mbf{v}_k$ bounded by zonotopes $W$ and $V$, respectively. Step 3 generates figures illustrating the resulting state $\mbf{x}_k$ and output $\mbf{y}_k$, whereas Steps 4 and 5 retrieve their respective values.

\begin{algorithm}[!tb] 
	\caption{Example of the capabilities of a \code{DTSystem} object with input matrices \code{A}, \code{Bw} and \code{C}, \code{Zonotope} objects \code{W} and \code{V}, and initial state \code{x0}, for \code{N} time steps}
	\label{alg:dtsystemexample}
	\scriptsize
	\begin{algorithmic}[1]
        \State \code{linearsys = DTSystem('linear','A',A,'Bw',Bw,'C',C);}
        \State \code{linearsys = simulate(linearsys,x0,N,W,V);}
        \State \code{plot(linearsys);}
        \State \code{x = linearsys.simdata.x;}
        \State \code{y = linearsys.simdata.y;}
	\end{algorithmic}
	\normalsize
\end{algorithm}

\section{Advanced features of {\acron}} \label{sec:advanced}

\subsection{State estimation of linear systems and descriptor systems} \label{sec:linearsystems}

Our library comes with algorithms implementing a few methods for state estimation of the linear systems \eqref{eq:systemlinear} and \eqref{eq:systemdescriptor}. These are located in the \code{estimation} folder. 

{\acron} implements state estimation of linear systems using zonotopes, constrained zonotopes, and line zonotopes. For bounded uncertainties $\mbf{w}_k \in W$ and $\mbf{v}_k \in V$, state estimation of linear systems consists of recursively computing enclosures $\bar{X}_k$ for the prediction step and $\hat{X}_k$ for the update step, given by, respectively,
\begin{align}
    \bar{X}_k \supseteq \{ & \mbf{A} \mbf{x}_{k-1} + \mbf{B}_w \mbf{w}_{k-1} + \mbf{B}_u \mbf{u}_{k-1} : \nonumber \\
    & \qquad (\mbf{x}_{k-1},\mbf{w}_{k-1}) \in \hat{X}_{k-1} \times W\}, \\
    \hat{X}_k \supseteq \{ & \mbf{C} \mbf{x}_k + \mbf{D}_v \mbf{v}_k = \mbf{y}_k : (\mbf{x}_{k},\mbf{v}_{k}) \in \bar{X}_k \times V \}.
\end{align}

The file \code{zon_linear_estimator.m} implements linear state estimation based on zonotopes, with prediction step given by 
\begin{equation} \label{eq:zonlinearpred}
    \bar{X}_k = \mbf{A} \hat{X}_{k-1} \oplus \mbf{B}_w W \oplus \mbf{B}_u \mbf{u}_{k-1},
\end{equation}
and update step given by
\begin{equation} \label{eq:zonlinearupdate}
    \hat{X}_k = \bar{X}_k \cap (\cap_{j=1}^{n_y} S_j(\mbf{y}_k)), 
\end{equation}
where $S_j(\mbf{y}_k)$ is a \code{Strip} object describing the set of states consistent with $\bar{X}_k$ and the $j$th element of $\mbf{y}_k$. Each strip $S_j(\mbf{y}_k)$ and the intersections in \eqref{eq:zonlinearupdate} are computed according to \cite{Bravo2006}.

The file \code{czon_linear_estimator.m} implements linear state estimation based on constrained zonotopes, as in \cite{Scott2016}. The prediction step is given by \eqref{eq:zonlinearpred}, with the update step computed as
\begin{equation} \label{eq:czonlinearupdate}
    \hat{X}_k = \bar{X}_k \cap_\mbf{C} (\mbf{y}_k \oplus (-\mbf{D}_v V)).
\end{equation}
Finally, the file \code{lzon_linear_estimator.m} implements linear state estimation based on line zonotopes, using \eqref{eq:zonlinearpred} for the prediction step and \eqref{eq:czonlinearupdate} for the update step. All the linear state estimation algorithms in {\acron} take a linear \code{DTSystem} object as input to facilitate the usage of the system matrices in the estimation algorithms.

{\acron} also implements state estimation of linear descriptor systems, using constrained zonotopes and line zonotopes. For bounded uncertainties $\mbf{w}_k \in W$ and $\mbf{v}_k \in V$, state estimation of \eqref{eq:systemdescriptor} consists in recursively computing enclosures $\bar{X}_k$ for the prediction step and $\hat{X}_k$ for the update step, given by, respectively,
\begin{align}
    \bar{X}_k \supseteq \{ & \mbf{x}_k \in \realset^{n_x} : \mbf{E} \mbf{x}_{k} = \mbf{A} \mbf{x}_{k-1} + \mbf{B}_w \mbf{w}_{k-1} + \mbf{B}_u \mbf{u}_{k-1}, \nonumber \\
    & \qquad (\mbf{x}_{k-1},\mbf{w}_{k-1}) \in \hat{X}_{k-1} \times W\}, \\
    \hat{X}_k \supseteq \{ & \mbf{C} \mbf{x}_k + \mbf{D}_v \mbf{v}_k = \mbf{y}_k : (\mbf{x}_{k},\mbf{v}_{k}) \in \bar{X}_k \times V \}.
\end{align}

The file \code{czon_descriptor_estimator.m} implements the state estimation method proposed in \cite{Rego2020b} for the computation of $\bar{X}_k$ and $\hat{X}_k$ using constrained zonotopes, while the file \code{lzon_descriptor_estimator.m} implements the state estimation method proposed in \cite{Rego2024LineZonArxiv} using line zonotopes. Both algorithms take a descriptor \code{DTSystem} object as input to facilitate the usage of the system matrices in the estimation algorithms.

\begin{remark} \rm \label{rem:faultdetection}
    The state estimation methods in {\acron} can be employed for fault detection, by verifying the inclusion of the measurement using \code{isinside} on the respective set.
\end{remark}

\subsection{Active fault diagnosis of linear systems} \label{sec:faultdiag}

The current version of {\acron} implements a method for active fault diagnosis of linear systems using zonotopes. This algorithm is located in the \code{faultdiag} folder. It addresses a system whose trajectories satisfy one of the possible models
\begin{subequations} \label{eq:systemlinearfault}
\begin{align}
    \mbf{x}_k & = \mbf{A}^{[i]} \mbf{x}_{k-1} + \mbf{B}_w^{[i]} \mbf{w}_{k-1} + \mbf{B}_u^{[i]} \mbf{u}_{k-1}, \\
    \mbf{y}_k^{[i]} & = \mbf{C}^{[i]} \mbf{x}_{k} + \mbf{D}_v^{[i]} \mbf{v}_{k}.
\end{align}
\end{subequations}

\begin{figure}[!tb]
	\centering{
		\def\svgwidth{0.95\columnwidth}
        {\tiny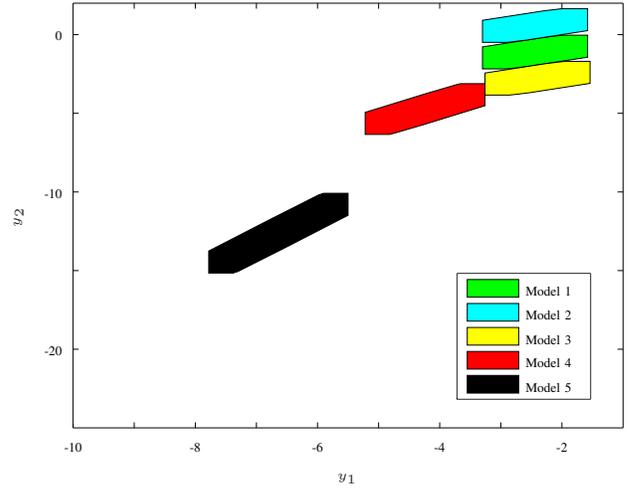}
		\caption{Separated output reachable tubes using the MIQP fault diagnosis approach proposed in \cite{Scott2014}, implemented in {\acron}. The \code{Zonotope} objects have been plotted using \code{plot}.}\label{fig:zonfaultdiagmiqp}}
\end{figure}

The file \code{zonAFD_inputdesign.m} implements the open-loop active fault diagnosis algorithm proposed in \cite{Scott2014}. It receives an array of linear \code{DTSystem} objects as input to describe \eqref{eq:systemlinearfault}, and allows the user to choose between solving a mixed-integer quadratic program (MIQP) or an analogous mixed-integer linear program (MILP), to design the input sequence that separates the output reachable tubes of the collection of models \eqref{eq:systemlinearfault}. An example is available in \code{demos/faultdiag}, which reproduces the first example from \cite{Scott2014} using the MIQP method (Fig.~\ref{fig:zonfaultdiagmiqp}), together with results obtained using an MILP approach.%

\subsection{Propagation of sets through nonlinear functions} \label{sec:propagnonlinear}

One of the main features of {\acron} is the implementation of algorithms for propagating sets through nonlinear functions. Such methods are essential for reachability analysis and state estimation of nonlinear systems and are available for \code{Interval}, \code{Zonotope}, and \code{CZonotope} objects. It follows a common syntax for the different objects, \code{propagate(X,fname,OPTS)}, in which: (i) \code{X} is an object of one of the three mentioned classes, (ii) \code{fname} is the name of the MATLAB function implementing the mathematical nonlinear function, and (iii) \code{OPTS} is an options structure allowing the user to choose which approximation method to use for propagation.%

For propagation of an \code{Interval} object, {\acron} implements the natural interval extension and the mean value extension \cite{Moore2009}. These methods are available in our library for completeness and for the possibility of comparison with zonotopic algorithms. For \code{Zonotope} objects, our library implements: (i) the mean value extension proposed in \cite{Alamo2005a}, (ii) the first-order Taylor extension proposed in \cite{Combastel2005}, and (iii) the DC programming approach described in \cite{Alamo2008}. In the current version of {\acron}, full support for DC decomposition is given only for the convexification method based on \cite{Adjiman1996aBB}.

Additionally, for \code{CZonotope} objects, our library implements: (i) the mean value and first-order Taylor extensions proposed in \cite{Rego2020} and \cite{Rego2021}, (ii) the DC programming approach developed in \cite{DePaula2024DC}, and (iii) the novel polyhedral relaxation approach proposed in \cite{Rego2024CZPRCDC} and \cite{Rego2025PolyRelaxArxiv}. The polyhedral relaxation approach relies heavily on the \code{Polyrelax} object implementation described in Section \ref{sec:polyrelax}, making use of its automated procedures for easy computation of the propagated constrained zonotope.

\begin{figure}[!tb]
	\centering{
		\def\svgwidth{\columnwidth}
        {\scriptsize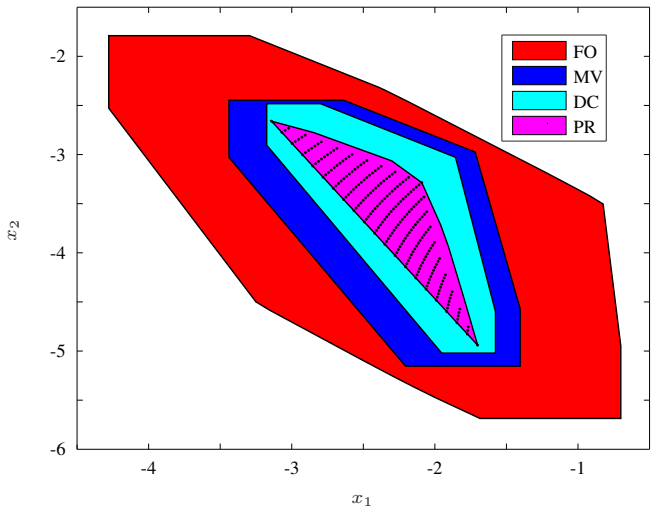}
		\caption{CZ enclosures of $\mbf{f}(X)$ obtained using {\acron}, along with propagated samples (black dots). The sets were plotted using \code{plot}.}\label{fig:czpropagate}}
\end{figure}

To illustrate the capabilities of {\acron} for propagation of sets through nonlinear functions, we present the example described in \code{demo_propagate.m}, located in \code{demos/basic/CZonotope/demo_propagate}. Let
\begin{equation*}
X \triangleq \left( \begin{bmatrix} 0.2 & 0.4 & 0.2 \\ 0.2 & 0 & -0.2 \end{bmatrix}, \begin{bmatrix} -1 \\ 1 \end{bmatrix}, 2{\cdot}\ones{1}{3}, -3\right)\czon,
\end{equation*}
and $\mbf{f}: \realset^2 \to \realset^2$ given by
\begin{align*}
    f_1(\mbf{x}) & \triangleq 3x_1 - \dfrac{x_1^2}{7} - \dfrac{4 x_1 x_2}{4 + x_1}, \\ 
    f_2(\mbf{x}) & \triangleq -2x_2 + \dfrac{3x_1x_2}{4+x_1}.
\end{align*}
This example consists on enclosing $\mbf{f}(X)$ by a constrained zonotope through the different approximation methods implemented in {\acron}: mean value extension (MV), first-order Taylor extension (FO), DC programming (DC), and polyhedral relaxations (PR). This is accomplished through \code{propagate(X,fname,OPTS)}. Fig.~\ref{fig:czpropagate} shows the obtained enclosures, along with propagated samples (black dots). The polyhedral relaxation approach provides the least conservative enclosure for this example.

\subsection{Nonlinear state estimation} \label{sec:estimationnonlinear}

Our library comes also with algorithms implementing a few methods for state estimation of nonlinear discrete-time systems as in \eqref{eq:systemnonlinear}. These are located in the \code{estimation} folder, and use \code{Zonotope} and \code{CZonotope} objects.

For bounded uncertainties $\mbf{w}_k \in W$ and $\mbf{v}_k \in V$, state estimation of \eqref{eq:systemnonlinear} consists in recursively computing enclosures $\bar{X}_k$ for the prediction step and $\hat{X}_k$ for the update step, given by, respectively,
\begin{align}
    \bar{X}_k \supseteq \{ & \mbf{f}(\mbf{x}_{k-1},\mbf{w}_{k-1},\mbf{u}_{k-1}) : \nonumber \\
    & \qquad \qquad (\mbf{x}_{k-1},\mbf{w}_{k-1}) \in \hat{X}_{k-1} \times W\}, \\
    \hat{X}_k \supseteq \{ & \mbf{x}_k \in \bar{X}_k : \mbf{g}(\mbf{x}_k,\mbf{v}_k) = \mbf{y}_k, \mbf{v}_{k} \in V \}.
\end{align}

The file \code{zon_meanvalue_estimator.m} implements a \code{Zonotope} state estimator with prediction step based on the mean value extension proposed in \cite{Alamo2005a}, while the file \code{zon_firstorder_estimator.m} implements the prediction step based on the first-order Taylor extension in \cite{Combastel2005}. The update step in both algorithms are given by the interval arithmetic approximation in \cite{Alamo2005a}, and intersections with strips by \cite{Bravo2006}.

The algorithm in \code{czon_meanvalue_estimator.m} implements a \code{CZonotope} state estimator with prediction and update steps based on the the mean value extension proposed in \cite{Rego2020,Rego2021}, while the algorithm in \code{czon_firstorder_estimator.m} implements both steps based on the first-order Taylor extension developed in the same references. Additionally, \code{czon_dcprog_estimator.m} implements the DC programming approach developed in \cite{DePaula2024DC}, while \code{czon_polyrelax_estimator.m} consists of the polyhedral relaxation approach proposed in \cite{Rego2025PolyRelaxArxiv}. 

\begin{figure}[!tb]
	\centering{
		\def\svgwidth{\columnwidth}
        {\scriptsize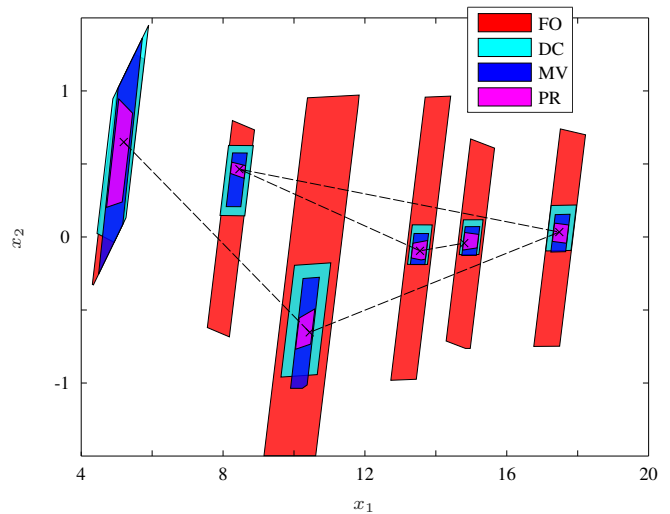}
		\caption{Estimated CZs $\hat{X}_k$ using different methods in {\acron}, along with $\mbf{x}_k$ ($\times$), for a few selected instants steps. The CZs were plotted using \code{plot}.}\label{fig:czestimationsets}}
\end{figure}

\begin{figure}[!tb]
	\centering{
		\def\svgwidth{\columnwidth}
        {\scriptsize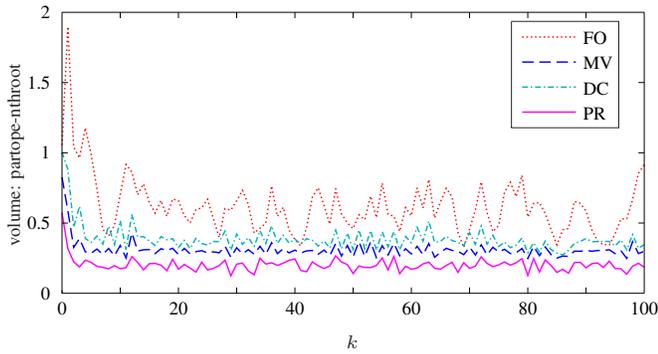}
		\caption{Approximated volume metric of the estimated CZs $\hat{X}_k$ using different methods in {\acron}.}\label{fig:czestimationvol}}
\end{figure}

Several examples are available in \code{demos/estimation}. One of these examples is illustrated here, which consists of state estimation of \eqref{eq:systemnonlinear} using CZs, through distinct approximation methods: mean value extension (MV), first-order Taylor extension (FO), DC programming (DC), and polyhedral relaxation (PR). Let
\begin{equation*}
X_0 \triangleq \left( \begin{bmatrix} 0.5 & 1 & -0.5 \\ 0.5 & 0.5 & 0 \end{bmatrix}, \begin{bmatrix} 5 \\ 0.5 \end{bmatrix}\right)\zon,
\end{equation*}
and $\mbf{f}: \realset^2 \times \realset^2 \to \realset^2$, $\mbf{g}: \realset^2 \times \realset^2 \to \realset^2$, given by
\begin{align*}
    f_1(\mbf{x},\mbf{w}) & \triangleq 3x_1 - \dfrac{x_1^2}{7} - \dfrac{4 x_1 x_2}{4 + x_1} + w_1, \\ 
    f_2(\mbf{x},\mbf{w}) & \triangleq -2x_2 + \dfrac{3x_1x_2}{4+x_1} + w_2, \\
    g_1(\mbf{x},\mbf{v}) & \triangleq x_1 - \sin\left(\frac{x_2}{2}\right) + v_1, \\
    g_2(\mbf{x},\mbf{v}) & \triangleq (-x_1 + 1)x_2 + v_2.
\end{align*}
The initial state is $\mbf{x}_0 = (5.2,0.65)$ and the uncertainties are bounded by $\|\mbf{w}_k\|_\infty \leq 0.5$ and $\|\mbf{v}_k\|_\infty \leq 0.2$. Fig.~\ref{fig:czestimationsets} shows the estimated CZs $\hat{X}_k$ for the different methods for a few time instants, along with the true trajectory $\hat{x}_k$. As in the propagation example, the polyhedral relaxations approach provides the least conservative enclosure for this example. This is further highlighted in Fig.~\ref{fig:czestimationvol}, which shows the approximate volume metric of $\hat{X}_k$ obtained using \code{volume(X,'partope-nthroot')}, for $k \in [0,100]$.

\section{Conclusions} \label{sec:conclusions}

This paper introduced a new MATLAB library for {\acrone} -- {\acron}, featuring user-friendly implementations of various set-based algorithms in the literature built upon zonotopic sets. Its capabilities include the basic set operations with various set representations, state estimation of linear and descriptor systems, linear active fault diagnosis, propagations of sets through nonlinear functions and nonlinear state estimation using various approximation methods from the literature. Efficient order reduction methods are also included for the implemented set representations. Some of the main functionalities of the new library were demonstrated in numerical examples. Future versions of {\acron} will include features not yet implemented or, with the advance of the theory, new developed ones.










\bibliography{\bibfolder/masterthesis_bib,\bibfolder/appendices_bib,\bibfolder/UAVControl_bib,\bibfolder/BackgroundHist_bib,\bibfolder/Surveys_bib,\bibfolder/PassiveFTC_bib,\bibfolder/ActiveFTC_bib,\bibfolder/UAVFTC,\bibfolder/SetTheoretic_bib,\bibfolder/SetTheoreticFTCFDI_bib,\bibfolder/Davide_bib,\bibfolder/paperAutomatica_bib,\bibfolder/paperCDC_bib,\bibfolder/paperECC_bib,\bibfolder/paperIFAC_bib,\bibfolder/paperNonlinearMeas_bib,\bibfolder/Robotic_bib,\bibfolder/stelios_bibliography,\bibfolder/phdthesis_bib,\bibfolder/paperParameter_bib,\bibfolder/Diego_bib,\bibfolder/paperMixed_bib,\bibfolder/paperPolyRelax_bib,\bibfolder/paperToolbox_bib}

\end{document}